\documentclass{aa}
\usepackage{graphics}
\begin{document}

\title{Magnetic Fields and Ionized Gas in the Local Group
Irregular Galaxies IC~10 and NGC~6822 }

\author{K.T. Chy\.zy\inst{1}
\and J. Knapik\inst{1}
\and D. J. Bomans \inst{2}
\and U. Klein\inst{3}
\and R. Beck\inst{4}
\and M. Soida\inst{1} 
\and M. Urbanik\inst{1}
}

\institute{Astronomical Observatory, Jagiellonian
University,
ul. Orla 171, 30-244 Krak\'ow, Poland
\and
Astronomisches Institut der Ruhr-Universit\"at Bochum,
Universit\"atsstr. 150, 44780 Bochum, Germany
\and
Radioastronomisches Institut der Universit\"at Bonn, Germany
\and
Max-Planck-Institut f\"ur Radioastronomie, Postfach 2024, 
D-53010 Bonn, Germany
}
\titlerunning {Magnetic Fields and Ionized Gas in
Irregular Galaxies}
\authorrunning{K. Chy\.zy et al.}
\mail{chris@oa.uj.edu.pl}
\offprints{K. Chy\.zy}
\date{Received 28 February 2003 / Accepted 24 April 2003}

\abstract{
We performed a high-sensitivity search for galaxy-scale
magnetic fields by radio polarimetry at 10.45~GHz and
4.85~GHz with the Effelsberg 100~m radio telescope,
accompanied by H$\alpha$ imaging, for the two Local Group
irregular galaxies IC~10 and NGC~6822. Their star-forming bodies are
small and rotate slowly. IC~10 is known to have a very high
star-forming activity, resembling blue compact dwarfs, while
NGC~6822 has a low overall star-formation level. Despite
very different current star formation rates, our H$\alpha$
imaging revealed a large web of diffuse H$\alpha$ filaments
and shells in both IC~10 and NGC~6822. Some of them extend
far away from the galaxy's main body. The total power
emission of both objects shows bright peaks either at the
positions of optically strong star-forming clumps (IC~10) or
individual \ion{H}{ii} regions or supernova remnants
(NGC~6822). However, in both cases we detect a smoothly
distributed, extended component. In IC~10 we found clear evidence 
for the presence of a diffuse, mostly random magnetic field 
of $\simeq14$\,$\mu$G strength, probably generated by a fluctuation dynamo.
One of the H$\alpha$-emitting filaments appears to be
associated with enhanced magnetic fields. We also rediscuss
the reddening of IC~10 and its implications for its distance.
In the case of NGC~6822 we found only very
weak evidence for nonthermal emission, except
perhaps for some regions associated with local gas
compression. We detect in both galaxies small spots of
polarized emission, indicative of regular fields ($\simeq 3$\,$\mu$G), 
at least
partly associated with local compressional phenomena.
}

\maketitle

\section{Introduction}

Irregular galaxies are low-mass objects exhibiting a variety
of rotational properties with a subclass of them rotating
very slowly (rotational speeds V$_\mathrm{rot}\le$30 km/s) and
often chaotically (e.g. Lo et al. \cite{lo93}). They constitute
important laboratories for large-scale interactions of stars
with the interstellar medium: the low gravitational
potential and relatively small size increase the probability
that superbubbles, forming close to star-forming regions,
may break out of the galaxy (e.g. Mac Low \&~Ferrara \cite{macl98}).
Many irregular galaxies exhibit giant arcs or filaments of
ionized gas (e.g. Sabbadin \&~Bianchini \cite{sabb79}, Hunter et al. 
\cite{hunt93}, Bomans et al. \cite{bom97}). The role of magnetic fields in
the origin and confinement of these ionized structures
(e.g. Hunter \& Gallagher \cite{hunt90}) is still a matter of debate.

In spiral galaxies magnetic fields, which are sufficiently strong to
trigger star formation via magnetic instabilities (Blitz
\&~Shu \cite{blitz80}) or to influence the superbubble expansion
(Ferriere et al. \cite{ferr91}), are probably generated by the 
mean-field dynamo (see Beck et al. \cite{beck96}). This requires strong
Coriolis forces (hence a rapid rotation) to give the
turbulent motions a preferred sense of twisting. A
sufficient size of the ionized gas envelope is also
required. Large irregulars with star-forming bodies of 8--10~kpc
in diameter still posses significant regular fields
(LMC -- Klein et al. \cite{klei93}, NGC~4449 -- Chy\.zy at al. \cite{chyz00}).
They are explicable by non-standard dynamos driven by Parker
instability (Moss et al. \cite{moss99}, Hanasz \& Lesch \cite{hana00}),
working at low rotation speeds. The efficiency of postulated
mechanisms is proportional to the ionized gas scale height,
which may make them inefficient in very small irregulars
having the star-forming bodies of 1--3~kpc in diameter.
Moreover, the magnetic field escape by diffusion is several
times faster in such small objects than in large irregulars.
Thus, small irregular galaxies should lack global magnetic
fields (Chy\.zy \cite{chyz02}). No relevant observational information on that matter
existed up to now.

Spiral galaxies show much more regular magnetic fields in
the interarm space than in strongly star-forming spiral arms
(e.g. Beck \& Hoernes \cite{BH96}). Nothing is known whether such a
difference also occurs between small very strongly and very
weakly star-forming irregulars. For all above reasons we
performed a sensitive search for extended, diffuse total
power and polarized emission in two small irregulars: IC~10
and NGC~6822 reflecting two extremities of star-forming
activity. The star-forming body of IC~10 has the diameter of
only 1.6~kpc assuming the distance of 0.8~Mpc (Wilson et al.
\cite{wils96}). Within this radius it rotates slowly at the speed
V$_\mathrm{rot}\simeq 30$~km/s (Wilcots \&~Miller \cite{wilc98}) and shows
two regions of a very high star-forming activity. The mean
H$\alpha$ surface brightness of this galaxy is 4 times
higher than in NGC~4449, making IC~10 a rapidly star-forming
object similar to young blue compact galaxies (BCG, Richer
et al. \cite{rich01}). The optical body of NGC~6822 is physically
about twice larger than IC~10. It lacks prominent regions of
recent star formation. Both galaxies posses similar \ion{H}{i} mass content:
$1.7\cdot10^8\,M_\odot$ in the case of IC~10 (Huchtmeier \cite{huch79}) 
and $1.3\cdot10^8\,M_\odot$ for NGC~6822 (de Blok \&~Walter
\cite{debl00}).

 As a search for weak extended emission is particularly
difficult for NGC~6822 because of its large angular size and
low radio surface brightness which makes single-dish
observations preferred. For the sake of comparison both
objects were studied in a way providing the best compromise
between sensitivity and resolution, ensuring a similar
number of beams per galaxy size. This was done by deep
mapping of NGC~6822 at 4.85~GHz and of IC~10 at 10.45~GHz
using the 100-m Effelsberg radio telescope.

To study the associations of magnetic fields with the
ionized gas and star formation we obtained sensitive images
of both objects in the H$\alpha$ line. The \ion{H}{ii} region
content of IC~10 was analyzed by Hodge \&~Lee (\cite{hodg90}) and
NGC~6822 by Hodge et al. (\cite{hodg88}) but in the latter case only
part of the galaxy was observed. Hunter et al. (\cite{hunt93}) found
that IC~10 is embedded in an impressive web of large
filaments and shells, while they could not detect similar
structures in NGC~6822. Our new H$\alpha$ data cover each
galaxy on one CCD frame with arcsecond resolution and high
sensitivity.
In Sect. 2 we present the observing techniques used to
obtain the maps of total power, polarized intensity and
H$\alpha$ emission. Section 3. presents the observational
results. The relative content of thermal and nonthermal
emission and the question of global magnetic fields is
discussed in Sect. 4.

\section{Observations and data reduction}

\subsection{Radio observations}

\begin{figure}
\resizebox{\hsize}{!}{\includegraphics{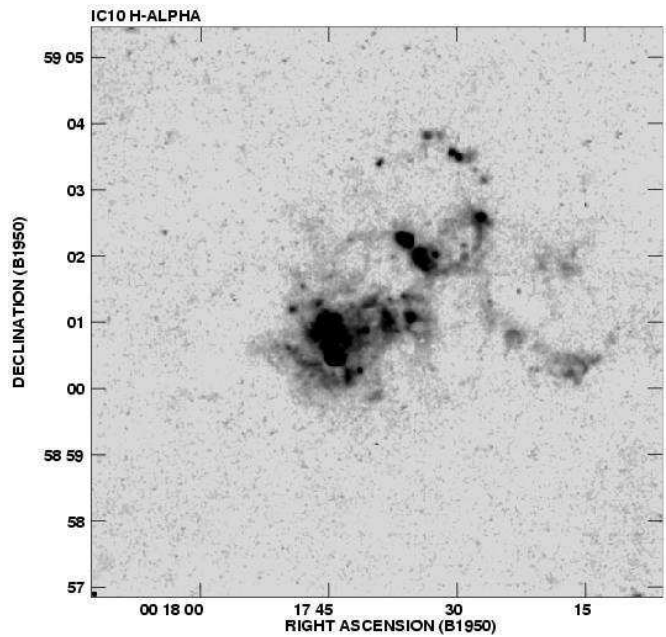}}
\caption{
 H$\alpha$ image of IC~10, free from the stellar continuum.
}
\label{icha}
\end{figure}

\begin{figure}
\resizebox{\hsize}{!}{\includegraphics{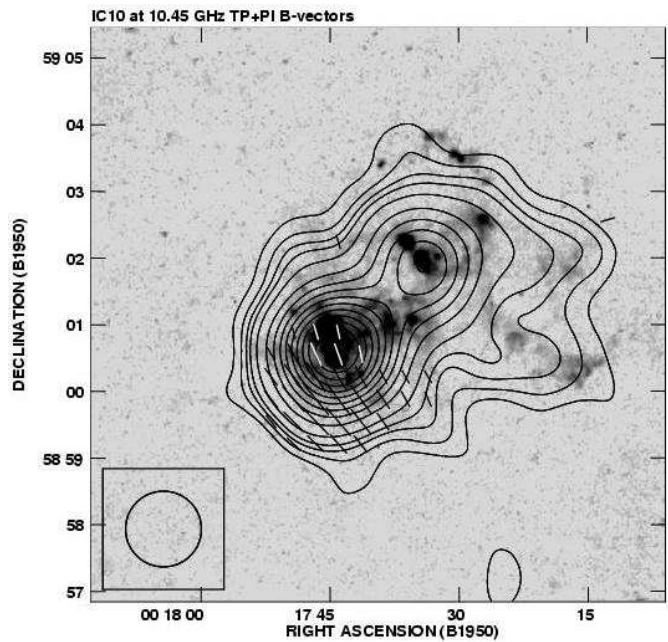}}
\caption{
Total power contour map of IC~10 at 10.45~GHz, with B-vectors
of polarized intensity superimposed onto the
H$\alpha$ image. The resolution is 1\rlap.\arcmin 13.
Contour levels are 2, 3, 4, 5 then 7, 11, 15 etc.~mJy/b.a.,
the vector of 1\arcmin \, length corresponds to polarized
intensity of 1.5~mJy/b.a.
 }
\label{ictp}
\end{figure}

The total power and polarization maps of IC~10 at 10.45~GHz
were obtained using the four-horn system in the secondary
focus of the Effelsberg 100-m MPIfR telescope (Schmidt
et~al. \cite{schm93})\footnote{The 100-m telescope at Effelsberg is
operated by the Max-Planck-Institut f\"ur Radioastronomie
(MPIfR) on behalf of the Max-Planck-Gesellschaft}, the
details of the observing procedure were described in detail
in our previous works (e.g. Soida et al. \cite{soi96}). In total we
obtained 28 coverages of IC~10 scanned in the 
azimuth-elevation frame. NGC~6822 was observed in total power and
polarization at 4.85~GHz using the two-horn system in the
secondary focus of the Effelsberg telescope. We obtained 9
azimuth-elevation coverages of this galaxy.

The telescope pointing was checked at time intervals of
about 2~hours by making cross-scans of nearby strong point
sources. The flux density scale was calibrated by mapping
the highly polarized source 3C286. Its total power flux
density of 4.47~Jy at 10.45~GHz and 7.44~Jy at 4.85~GHz has
been adopted using the formulae by Baars et~al. (\cite{baa77}). The
same calibration factors were used for total power and
polarized intensity, yielding a mean degree of polarization
of 3C286 of 12.2\% at 10.45~GHz and 10.5\% at 4.85~GHz, in
reasonable agreement with other published values (Tabara
\&~Inoue \cite{taba80}).

The data reduction was performed using the NOD2 data
reduction package (Haslam \cite{hasl74}). By combining the
information from appropriate horns, using the ``software
beam-switching'' technique (Morsi \&~Reich \cite{mors86}) followed by
a restoration of total intensities (Emerson et~al. \cite{emer79}),
for each coverage we obtained I, Q and U maps. These were
combined into final maps of total power, polarized
intensity, polarization degree and polarization position
angles using the spatial frequency weighting method (Emerson
\&~Gr\"ave \cite{emer88}). A digital filtering process, which removes
spatial frequencies corresponding to noisy structures
smaller than the telescope beam, was applied to the final
maps.

The original beam of our 10.45~GHz observations was 1\farcm
13, corresponding to 270 pc at the distance to IC~10 of 820
kpc (Wilson et~al. \cite{wils96}). At the distance to NGC~6822 of 490
kpc (McGonegal et~al. \cite{mcgon83}) the original beam of 2\farcm 5 at
4.85~GHz corresponds to 360 pc. To increase the sensitivity
to extended structures we also used data convolved to
1\farcm 3 (IC~10) and $3\arcmin$ (NGC~6822).

\subsection{Observations in H$\alpha$ line}

\subsubsection{IC~10}

We observed IC~10 using the Calar Alto Observatory 1.23~m
telescope in October 1994. A Tektronics $1024^2$ pixel CCD
was used as the detector resulting in a spatial scale of
0\farcs 502 per pixel. The seeing during our observations
was 1\farcs 3 and the conditions were non-photometric, with
light cirrus suspected. We took two 30 min exposures through
a 75\AA{} wide filter centered at 6560\AA, well matched to
the H$\alpha$ line of IC~10 (V$_\mathrm{rad} = -360$ km/s).
For the continuum image we observed IC~10 with a standard
Johnson-Cousins R filter (10 min). The data reduction was
performed in the standard way using the IRAF package. We
applied small kernal Gaussian filters to both the H$\alpha$
and the R image to match the slightly different point spread
functions. After this step the continuum subtraction was
performed, as described in e.g. Bomans et al. (\cite{bom97}). Due to
the poor photometric conditions we performed the flux
calibration relative to the H$\alpha$ maps of Hodge \&~Lee
(\cite{hodg90}). The major uncertainty of this process is the choice
of the aperture and centering. The scatter of the
correlation implies that our calibration has an uncertainty
of 25\%. The correction of our H$\alpha$ fluxes for the
contribution of [\ion{N}{ii}] is less than 1.2 for \ion{H}{ii} regions. The
contribution may be higher for the filaments, due to the
larger [\ion{N}{ii}]/H$\alpha$ ratio in the diffuse ionized gas
(DIG) (see e.g. Dettmar \cite{dett93}).

\subsubsection{NGC~6822}

The optical data for NGC~6822 were taken with the 1m telescope
at Mt. Laguna Observatory using a Loral/lesser $2048^2$ pixel CCD
giving a scale of 0.41\arcsec/pix. The data reduction techniques
were the same as for the observations of IC~10.
Seeing was around 1\farcs5. We observed NGC~6822 twice for 1800\,s
in an H$\alpha$ filter and for 600s in a broad band R filter.
The photometric conditions during observation of NGC~6822 were
unstable during the run. We therefore decided to calibrate our data
relatively to the published \ion{H}{ii} region fluxes by Hodge \&~Lee
(\cite{hodg90}).
The quality of the correlation is similar to that for IC~10,
as is also the correction of the flux scale for
contribution of [\ion{N}{ii}].

\section{Results}

\subsection{H$\alpha$ image of IC~10}

Our H$\alpha$ image of IC~10 (Fig.~\ref{icha}) shows the \ion{H}{ii}
region complexes in the SE and NW of the galaxy as well as
the extended diffuse ionized medium of the galaxy. Many of
the features were already described by Hunter et al. (\cite{hunt93}).
We point to a few features important for further discussion:
Most of the diffuse emission is actually in filaments and
shells, there is no evidence for an unresolved component.
The filaments and shells extend out to 0.6~kpc from the main
body of IC~10, and not all of them originate in the two
dominant star forming regions. Two very prominent H$\alpha$-emitting 
features extend westwards from the northern 
star-forming region. We also detect a faint bubble-shaped
H$\alpha$ feature SW of the prominent dust lane in the
southern part of IC~10, coincident with the position of the
non-thermal superbubble detected there by Yang \&~Skillman
(\cite{yang93}).

\subsection{Radio continuum of IC~10}

The 10.45~GHz total power map of IC~10 is shown in
Fig.~\ref{ictp} at full resolution with superimposed B-vectors 
of polarized intensity. The map has an r.m.s. noise
of 0.6~mJy/b.a. Two bright, poorly resolved total power
peaks are found at the positions of optically bright star-forming 
regions as already stated by Klein et al. (\cite{klei83}).
They are present in Condon's (\cite{cond87}) map at 1.49~GHz, thus
contain some nonthermal component, in agreement with earlier
findings by Klein \& Gr\"ave (\cite{klei86}). In addition IC~10 shows
smooth, extended emission (visible in Condon's map, too)
extending westwards up to 0.7~kpc from the star formation
nests. In the western disk, the diffuse total power emission
makes two extensions coincident with the H$\alpha$-emitting
filaments mentioned in the previous section. This region
coincides also with a faint, diffuse optical glow.
Integration of the total power map in elliptical rings with
an inclination of 33.6$\degr$ and position angle of
136$\degr$, both taken from the Lyon-Meudon Extragalactic
Database (LEDA), yields an integrated flux density of IC~10
at 10.45~GHz of $155\pm16$ mJy.

\begin{figure}
\resizebox{\hsize}{!}{\includegraphics{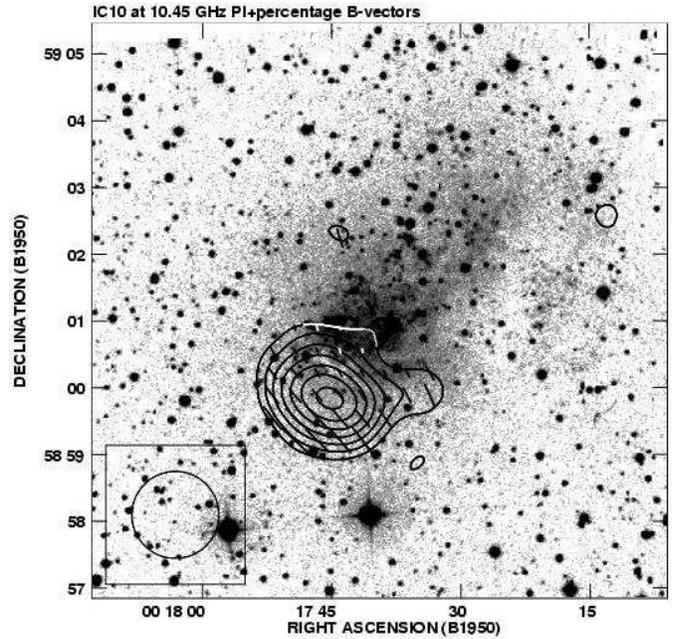}}
\caption{
Contour map of the polarized intensity of IC~10 at
10.45~GHz, with
B-vectors of the polarization degree superimposed onto an
optical image
from DSS. The resolution is 1\rlap.\arcmin 3. The contour
levels are
0.3, 0.4 etc.~mJy/b.a., the vector of 1\arcmin \, length
corresponds to polarization degree of 15\%
}
\label{icpi}
\end{figure}

Our pola\-rized inten\-sity map of IC~10 (Fig.~\ref{icpi})
has an r.m.s. noise of 0.14~mJy/b.a.. The polarized
brightness shows a single, weakly resolved peak south of a
straight, linear dust feature crossing the outskirts of the
southern star-forming region. The degree of polarization at
the polarized intensity peak is about 4\% increasing to 
8--10\% at its southern boundary. The orientation of the 
B-vectors is parallel to the dust lane. IC 10 lies close to
the galactic plane, where the Faraday rotation measures may
reach 200~rad/m$^2$ (Simard-Normandin \& Kronberg \cite{sim80}). At
the frequency of 10.45~GHz this corresponds to a
polarization angle offset by 9\degr, which does not affect
strongly the alignment with the dust lane. The bright total
power peaks are found to be completely unpolarized, with
polarization degrees at their position smaller than 0.7\%.
The diffuse emission in the northern disk and the western
extension are polarized by less than 0.8\%.

The polarized feature is not associated with any discrete
sources visible in Condon's (\cite{cond87}) map and is unlikely to be
due to a background source. Instead, it coincides with a
diffuse nonthermal spur found there by Yang \& Skillman
(\cite{yang93}). Integration of the polarized intensity map shown in
Fig.~\ref{icpi} in the same rings as the total power one
yields an integrated polarized flux density of IC~10 at
10.45~GHz of $1.9\pm1.1$~mJy. This implies a mean
polarization degree of $1.2\pm0.7$\%, much lower than in
normal spirals (Knapik et al. \cite{knap00}). All the detected
polarized flux comes from the barely resolved blob in the
southern region. After its subtraction as an unresolved
source, the polarization degree drops below 0.5\%.


\begin{figure}
\resizebox{\hsize}{!}{\includegraphics{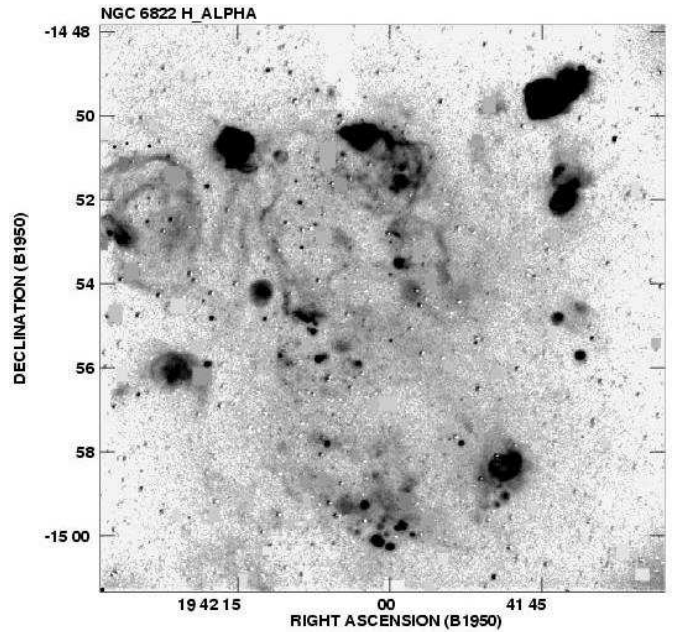}}
\caption{
H$\alpha$ image of NGC~6822, free from the contribution from
the
stellar continuum
}
\label{68ha}
\end{figure}

\begin{figure}
\resizebox{\hsize}{!}{\includegraphics{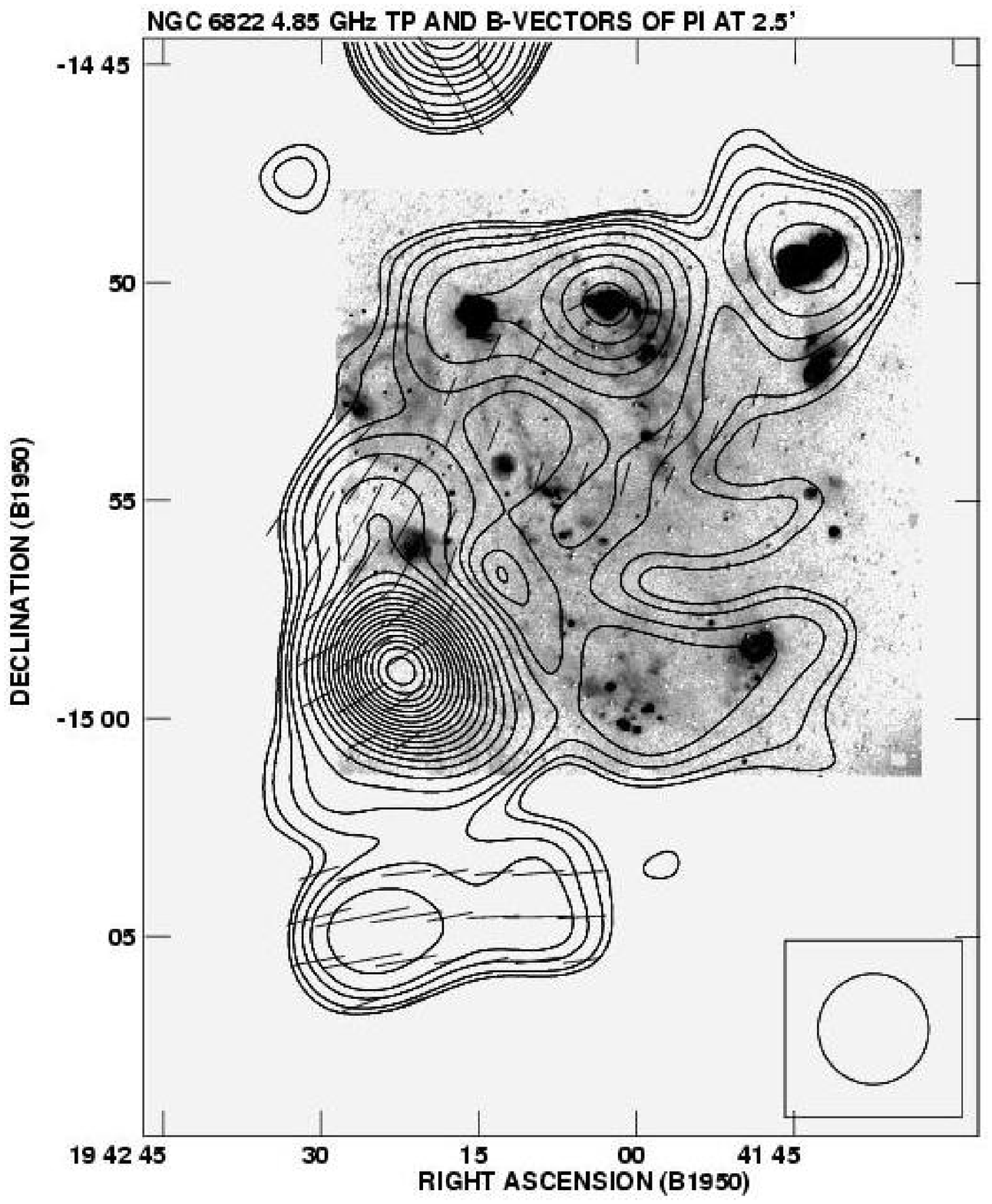}}
\caption{
Total power contour map of NGC~6822 at 4.85~GHz with B-vectors of
polarized intensity superimposed onto a H$\alpha$ image. The resolution
is 2\farcm 5. Contour levels are 1.5, 2, 3, 4, then 6, 9, 12
etc.~mJy/b.a., the vector of 1\arcmin \, length corresponds
to polarized intensity of 0.5~mJy/b.a.
}
\label{68tp}
\end{figure}

\subsection{H$\alpha$ image of NGC~6822}

Our new continuum-free H$\alpha$ image of NGC~6822
(Fig.~\ref{68ha}) covers only the central part of the
galaxy. Its detailed discussion will be presented by Bomans
 (in prep.). It shows small \ion{H}{ii} regions at several
positions in the galaxy's body. Some \ion{H}{ii} regions beyond the
radius of 1 kpc are seen for the first time. The size and
luminosity of the \ion{H}{ii} regions outside the main galaxy's body
are larger than those inside it. No large, dominant 
star-forming complexes like those in IC~10 were found. Giant \ion{H}{ii}
regions are missing in NGC~6822, too.

Our data confirm the large difference in current star
formation between IC~10 and NGC~6822, the former one having
more \ion{H}{ii} regions and they are more luminous than in
NGC~6822. The star-forming regions of NGC~6822 are
distributed over a much larger surface area than those in
IC~10. The number of superbubbles is also larger in NGC~6822
than in IC~10. In the latter the \ion{H}{ii} region population is
dominated by compact \ion{H}{ii} regions, occasionally accompanied
by high surface brightness arcs.

Surprisingly, despite its very low current star formation
rate NGC~6822 shows an extended network of diffuse H$\alpha$
filaments and shells, seen for the first time in the present
work. They are filling almost the whole optically bright
galaxy. Some of them extend out to 1.5~kpc from the main
body, especially to the NE.

\subsection{Radio continuum of NGC~6822}

Our total power map of NGC~6822 at 4.85~GHz
(Fig.~\ref{68tp}) has an r.m.s. noise of 0.65~mJy/b.a. Some
individual point sources coincide with isolated \ion{H}{ii} regions
and supernova remnants. However, the brightest peak east of
the optically bright galaxy and the double source in its
southern part have no obvious optical counterparts. In
addition to these discrete sources, seen also in the
1.49~GHz map by Condon (\cite{cond87}), we detect diffuse emission
extending westward from the brightest peak over the
optically bright body and filling the gaps between
individual sources. The diffuse radio emission west of the
strongest radio peak was also marginally detected by Klein
\& Gr\"ave (\cite{klei96}), while in our map we could measure it at a
level of at least $5\sigma$ r.m.s. noise (see Sect. 4).
Integration of the total power map in elliptical rings with
an inclination of 25$\degr$ and a position angle of 5$\degr$
(both taken from the LEDA database) yield an integrated flux
density of NGC~6822 at 4.85~GHz of $147\pm19$ mJy. The
background-like strong source located at R.A.$_{1950}=19^\mathrm{h}
42^\mathrm{m} 22\fs 9$ Dec$_{1950}=-14\degr 58\arcmin 43\arcsec$
with a flux of 48~mJy (see Section. 4.1.3) was subtracted in this derivation.

\begin{figure}
\resizebox{\hsize}{!}{\includegraphics{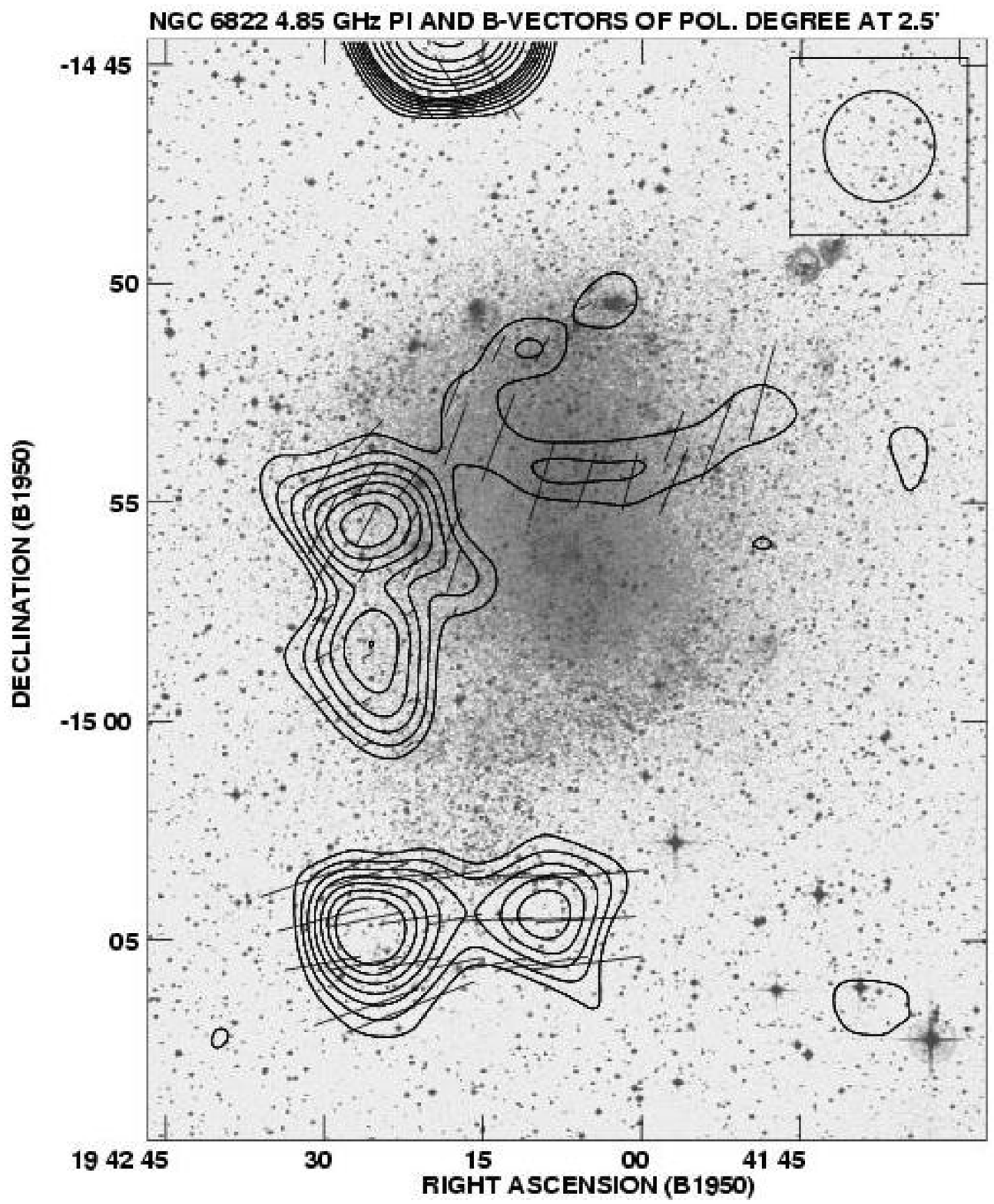}}
\caption{
Contour map of the polarized intensity of NGC~6822 at
4.85~GHz with B-vectors of the polarization degree
superimposed onto an optical image from DSS. The resolution 
is 2\farcm 5. Contour levels are 0.3,
0.4, 0.5 etc.~mJy/b.a., the vector of 1\arcmin \, length
corresponds to polarization degree of 8\%
}
\label{68pi}
\end{figure}

Our map of the polarized intensity of NGC~6822 at 4.85~GHz
(Fig.~\ref{68pi}) has an
r.m.s. noise of 0.1~mJy/b.a.. The bright total power source
east of the optical galaxy (a background object?) is weakly
polarized, with a degree of polarization of about 2\%.
Significant polarization was detected to extend northwards
of it, with a degree of polarization of about 8\%.
Significant polarization (mean degree of about 13\%) was
detected in the double structure (also background sources?)
in the southern galaxy's outskirts. Integration of the map
in the same elliptical rings as the total power one yields
an integrated polarized flux density of NGC~6822 at 4.85~GHz
of $5.1\pm2.1$~mJy implying a mean polarization degree at
this frequency of $3.5\pm1.5$\%, most of it being due to the
mentioned discrete sources. However, the diffuse radio
emission filling the inner galaxy (excluding possible
background sources and those related to star-forming regions
in the north) shows a mean polarization degree of
$5\pm1$\%. The degree of diffuse polarization is highest in
the rectangular region delineated by R.A.$_{1950}=
19^\mathrm{h}41^\mathrm{m}40^\mathrm{s}$ and
$19^\mathrm{h}42^\mathrm{m}13^\mathrm{s}$,
Dec$_{1950}=-14\degr57\arcmin$ and $-14\degr52\arcmin$ and reaches
$15\pm2$\%, extending over an area of about 6 beam areas.


\section{Discussion}

\subsection{Thermal and nonthermal emission }

\subsubsection{Thermal emission from the rapidly 
star-forming IC~10}

\begin{table}[t]
\caption{Point sources subtracted from IC 10 maps}
\label{points}
\begin{flushleft}
\begin{tabular}{ccrr}
\hline\noalign{\smallskip}
R.A.$_{1950}$& Dec$_{1950}$ & S$_{1.49}$ & S$_{10.45}$ \\
 & & $[$mJy$]$&$[$mJy$]$ \\
\hline\noalign{\smallskip}
$00^\mathrm{h} 17^\mathrm{m} 44\fs1$ & $59\degr 00\arcmin 28\arcsec $ & 3.6 &
3.0 \\
$00^\mathrm{h} 17^\mathrm{m} 44\fs9$ & $59\degr 00\arcmin 31\arcsec$ & 2.2 &
1.7 \\
$00^\mathrm{h} 17^\mathrm{m} 43\fs9$ & $59\degr 00\arcmin 51\arcsec$ & 3.8 &
4.2 \\
$00^\mathrm{h} 17^\mathrm{m} 44\fs6$ & $59\degr 01\arcmin 01\arcsec$ & 2.6 &
1.8 \\
$00^\mathrm{h} 17^\mathrm{m} 34\fs4$ & $59\degr 02\arcmin 02\arcsec$ & 10.4 &
8.4 \\
$00^\mathrm{h} 17^\mathrm{m} 36\fs2$ & $59\degr 02\arcmin 17\arcsec$ & 2.0 &
1.3 \\
$00^\mathrm{h} 17^\mathrm{m} 31\fs9$ & $59\degr 02\arcmin 16\arcsec$ & 2.4 &
0.6 \\
\hline\noalign{\smallskip}
\end{tabular}
\end{flushleft}
\end{table}

\begin{figure}
\resizebox{\hsize}{!}{\includegraphics{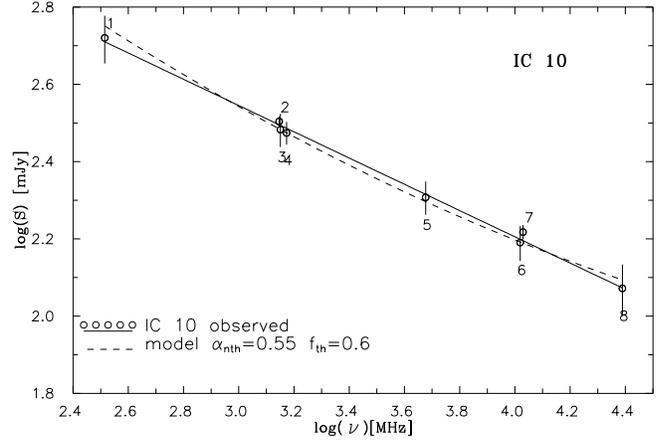}}
\caption{
The radio spectrum of IC 10. The numbers label the data
sources: 1. Computed by us from the WENSS survey maps
(Rengelink et al. \cite{reng97}), 2. White \& Becker (\cite{white92}), 3.
Shostak (\cite{shos74}), 4. Condon (\cite{cond87}),
5. Klein et al. (\cite{klei83}), 6.
this work, 7. Klein et al. (\cite{klei83}), 8. Klein \& Gr\"ave
(\cite{klei86}). The dashed line shows the best model spectrum
resulting from a combination of maximum thermal fraction of
60\% and a nonthermal spectral index of 0.55.
}
\label{spec}
\end{figure}

The radio spectrum of IC 10 between 1.4~GHz and 24~GHz is
presented in Fig.~\ref{spec}. The mean spectral index is
$\alpha=0.33 \pm 0.02$ (S$_{\nu}\propto \nu^{-\alpha}$).
This is much less than the source spectrum of cosmic-ray
(CR) electrons accelerated in shocks, yielding $\alpha=0.5$.
The spectrum observed in normal galaxies, involving
synchrotron losses and energy-dependent diffusion (Niklas et
al. \cite{nikl97}) is even steeper. Our attempts to model a
combination of thermal and nonthermal components (the latter
with an adjustable slope) yields the combination of a
nonthermal spectral index $\alpha_\mathrm{nt} = 0.55\pm 0.05$ and a
mean thermal fraction $\overline{f_\mathrm{th}} = 0.6\pm 0.1$ as
the best reproduction of the data (Fig.~\ref{spec}).
Nevertheless, higher $\overline{f_\mathrm{th}}$ reaching 0.7--0.75
corresponding to $\alpha_\mathrm{nt}\simeq$ 0.66--0.74 cannot
be excluded on the basis of radio spectrum alone.

We note that the radio spectrum of IC~10 is considerably
flatter than that of the much larger irregular NGC~4449, the
latter having an observed spectral slope of 0.41$\pm0.01$
(Klein et al. \cite{klei96}). These authors determined $\alpha_\mathrm{nt} =
0.7\pm 0.14$ and $\overline{f_\mathrm{th}}$ (converted to
10.45~GHz) of 0.4. The large irregular NGC~4449 seems to be
an intermediate case between large spirals and the small
irregular galaxy IC~10 concerning both thermal fraction and
nonthermal spectrum slope.

\begin{figure}
\resizebox{\hsize}{!}{\includegraphics{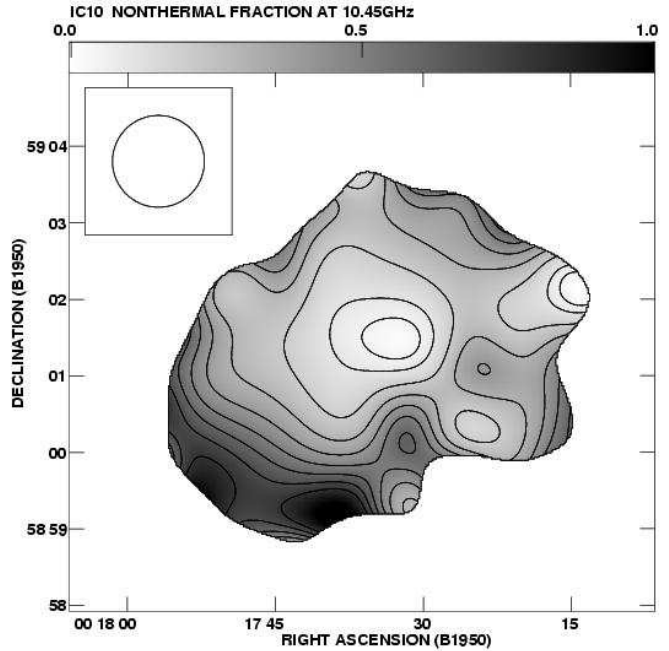}}
\caption{
Distribution of the nonthermal fraction in IC~10 computed
from the spectral index map between 1.49~GHz (Condon \cite{cond87})
and our map shown in Fig.~\ref{ictp}, assuming a nonthermal spectral
index of 0.55. The point sources listed in Table~\ref{points} were
subtracted}
\label{fth}
\end{figure}

Being aware of uncertainties due to extinction we verified
our estimates of the thermal flux using the H$\alpha$ map
convolved to the resolution of 72$\arcsec$. To determine the
extinction we compared the radio fluxes of thermal discrete
sources in the 6\,cm map of Yang \& Skillman (\cite{yang93}) to the
predictions from the H$\alpha$ emission of associated
\ion{H}{ii} regions. As a mean value we obtained
A$_\mathrm{H\alpha}$ = 4.35. We note however, that the metallicity
of IC 10 is several times smaller than that of our Galaxy
(Lequeux et al. \cite{leq79}), thus most of extinction
originates in the Galactic foreground, especially at the
Galactic latitude of $-$3.3\degr(LEDA database). As small
variations of Galactic extinction over the angular size of
IC 10 is expected this value has been applied to the whole
map. The results were compared to those obtained from the
radio spectrum. The difference at 10.45~GHz reaches
6~mJy/b.a. in the northern blob and 4~mJy/b.a. in the
southern one, thus on average it constitutes less than 30\%
of the thermal brightness. This is the maximum uncertainty
of all assumptions used in determining the thermal flux
(including constant extinction across the galaxy). In the
outer disk we find good agreement between the thermal
brightness estimated from the radio spectrum and the
H$\alpha$ emission. As the determination of the thermal flux
from the H$\alpha$ brightness was independent of our
assumption concerning $\alpha_\mathrm{nt}$, this supports in
retrospect our adopted values of $\alpha_\mathrm{nt}$=0.55 and
$\overline{f_\mathrm{th}}$=0.6.

Our estimate of the extinction results in E$_\mathrm{(B-V)}$ =
1.74, assuming a Mathis (\cite{math79}) reddening curve. This is very
similar to what was derived by Yang \& Skillman (\cite{yang93}) using
the same method, but a magnitude higher than the value
obtained from an analysis of the colour-magnitude diagram
(CMD) of IC10, E$_\mathrm{(B-V)}$ = 0.77 (Hunter \cite{hunt00}, Massey \&
Armandroff \cite{mass95}). Since a global high value of the reddening
seems incompatible with the CMD analysis (Hunter \cite{hunt00}), we
are forced to propose that the \ion{H}{ii} regions show significant
internal reddening, and/or there are pockets of high
foreground reddening. The first possibility is unlikely,
taking into account a reasonable agreement between our
thermal fluxes derived from the radio spectrum (insensitive
to extinction) and the H$\alpha$ line, assuming the same
extinction in \ion{H}{ii} regions and diffuse gas. Still, it is
interesting to note here, that the foreground extinction
maps of Schlegel et al. (\cite{schl98}) give a value for the
foreground reddening alone which is very similar to our
value. Clearly the debate about the reddening of IC10 is
still not settled yet. Since the reddening has implications
for the distance and the interpretation of the fluxes new
approaches could really help here. It is important to use
the Balmer decrement method for \ion{H}{ii} regions and the
widespread diffuse emission to derive a fine gridded
reddening map of IC 10.

\subsubsection{Nonthermal emission from IC 10}
\label{ic-nth}

Using the spectral index distribution computed between our
and Condon's (\cite{cond87}) map at 1.49~GHz and assuming a constant
nonthermal spectral index of 0.55 over the whole galaxy, we
have obtained the distribution of thermal and nonthermal
fractions in IC~10 at 10.45~GHz (the latter shown in
Fig.\ref{fth}). We computed the 10.45~GHz fluxes of point
sources present in the maps of Yang \& Skillman (\cite{yang93}, Table~\ref{points})
and subtracted them from our total power map, thus the
Figure refers to diffuse emission only. The nonthermal
fraction shows a minimum at the positions of the 
north-western bright star-forming region, dropping to zero there
independently of the assumed $\alpha_\mathrm{nt}$. The nonthermal
emission associated with the southern star-forming region is
larger and amounts to 35\%. If a steeper nonthermal spectrum
is assumed, e.g. $\alpha_\mathrm{nt} = 0.7$ this value changes to
20\%. The nonthermal fraction rises gradually towards the
southern galaxy boundary, reaching some 70\%--90\% ($\simeq$
45\%--55\% for $\alpha_\mathrm{nt}$ of 0.7) in the nonthermal
blob observed by Yang \& Skillman (\cite{yang93}). We state the
existence of significant magnetic fields (partly regular) in
this region. Very little nonthermal emission is observed
from the northern H$\alpha$-emitting clump on the western
galaxy side, where the total power extension is almost
purely thermal. A long filament extending from IC~10 to the
SW is accompanied by a ridge with a nonthermal fraction of
about 40\% (assuming $\alpha_\mathrm{nt}=0.55$).

In IC~10, which is forming stars more intensively than
NGC~6822 we detected unambiguously a nonthermal component of
its radio continuum emission (amounting to 30\%--40\%) of
the total flux, indicating a widespread magnetic field. Its
mean equipartition strength within the area delineated by
5\% of the maximum brightness (after the subtraction of
sources from Yang \& Skillmann \cite{yang93}, Table~\ref{points}) is $14\pm
4$\,$\mu$G, changing only little whether we assume either a
thermal fraction of 60\% and a nonthermal spectral index of
0.55 or 70\% thermal fraction and a nonthermal spectral
index of 0.65. We adopted a lower energy cutoff of 300~MeV,
a proton-to electron ratio of 100 (Pacholczyk \cite{pach70}) and a
disk thickness of 0.8~kpc (assuming that IC~10 is roughly a
prolate ellipsoid). The error of the magnetic field strength
includes the uncertainty of these quantities by a factor of
2. The minimum field strength obtained by varying all
assumptions is 10\,$\mu$G. In the disk outskirts we obtain a
total equipartition magnetic field of 7\,$\mu$G. These values
are comparable to total equipartition fields in massive and
strongly star-forming spirals. Apparently the small size of
IC~10 does not prevent it from developing strong total
(mostly random) magnetic field. The nonthermal emission and
the total magnetic field thus fills the entire body of
IC~10, dominating in its southern region.

The best agreement between the thermal emission of IC~10
determined from the radio spectrum and H$\alpha$ emission
implies $\alpha_\mathrm{nt}$=0.55, hence a CR electron energy
spectrum even harder than in NGC~4449 (Klein et al, \cite{klei96}).
In fact, apart from the western extension the total power
emission drops almost to zero at a distance of $\simeq$ 400--500\,pc
from the large star-forming complexes, compared to
a typical scale length of $\simeq$1\,kpc in normal spirals.
This may mean the lack of an extended magnetic halo and no
space where the electrons could lose efficiently their
energy. The lack of an extended halo also allows fast escape
of cosmic rays -- they seem to be only weakly
confined by the magnetic field. The same conclusion has been
reached by Klein et al. (\cite{klei91}) for a larger sample of BCG's,
which show a clear radio spectrum steepening with increasing
luminosity, thus also with the mass of the stellar body. The
thermal content seems to increase and the spectrum of CR
electrons gets harder (as they are weaker confined) in a
continuous way from massive spirals, through large
irregulars to small, compact rapidly star-forming galaxies.

\subsubsection{Any nonthermal emission from the weakly star-forming NGC~6822?}

The analysis of the radio spectrum of NGC~6822 is difficult
because of the large angular extent and the too low level of
emission. Attempts to construct a spectral index map
between 4.85~GHz and 1.49~GHz led to unphysical results
because of severe flux losses at the latter frequency, even
in the vicinity of bright peaks. Instead, we attempted to
determine the flux densities at 4.85~GHz of particular point
sources identifiable in Condon's (\cite{cond87}) map at 1.49~GHz,
using the ``best subtraction'' method. This was done by
subtracting at the given position the image of a point
source with the flux density adjusted to reach a smoothly
distributed emission in its environment. This procedure
yielded some guess of the flux densities of subtracted
sources at 4.85~GHz which together with Condon's data
allowed us to determine their spectral indices between these
frequencies. All uncertainties of this procedure (determined
by varying the 4.85~GHz fluxes of subtracted sources) are
included in the spectral index errors. The results are
summarized in Table~\ref{spectral}.

\begin{table}[t]
\caption{Spectral index of discrete sources subtracted from
the map of NGC~6822}
\label{spectral}
\begin{flushleft}
\begin{tabular}{llrr}
\hline\noalign{\smallskip}

 R.A. $_{1950}$ & Dec$_{1950}$ & S$_{4.85}$& spectral \\
 & & mJy & index \\
\hline\noalign{\smallskip}
$19^\mathrm{h} 41^\mathrm{m} 51\fs1$ & $-14\degr 48\arcmin 00\arcsec$ &
$3.0\pm 0.7$
 & $0.86\pm 0.23$ \\
$19^\mathrm{h} 41^\mathrm{m} 43\fs3$ & $-14\degr 49\arcmin 27\arcsec$ &
$9.8\pm 0.8$
 & $0.14\pm 0.10$ \\
$19^\mathrm{h} 42^\mathrm{m} 15\fs5$ & $-14\degr 50\arcmin 38\arcsec$ &
$7.4\pm 0.8$
 & $0.16\pm 0.12$ \\
$19^\mathrm{h} 42^\mathrm{m} 02\fs9$ & $-14\degr 50\arcmin 30\arcsec$ &
$17.9\pm 0.7$
 & $0.11\pm 0.06$ \\
$19^\mathrm{h} 42^\mathrm{m} 25\fs0$ & $-14\degr 55\arcmin 36\arcsec$ &
$11.2\pm 0.8$
 & $0.79\pm 0.07$ \\
$19^\mathrm{h} 42^\mathrm{m} 22\fs9$ & $-14\degr 58\arcmin 43\arcsec$ &
$48.4\pm 2.0$
 & $0.96\pm 0.04$ \\
$19^\mathrm{h} 42^\mathrm{m} 17\fs8$ & $-14\degr 59\arcmin 49\arcsec$ &
$16.4\pm 1.5$
 & $0.57\pm 0.08$ \\
$19^\mathrm{h} 42^\mathrm{m} 24\fs9$ & $-15\degr 04\arcmin 51\arcsec$ &
$8.2\pm 1.0$
 & $0.65\pm 0.12$ \\
$19^\mathrm{h} 42^\mathrm{m} 09\fs3$ & $-15\degr 04\arcmin 24\arcsec$ &
$4.2\pm 0.7$
 & $0.67\pm 0.16$ \\
\noalign{\smallskip}
\hline
\end{tabular}
\end{flushleft}
\end{table}


The brightest total power peak (Fig.\ref{68tp}) east of the optically bright
galaxy is largely nonthermal. As it is partly polarized (Fig.\ref{68pi}) it
is likely to be a background source unrelated to the galaxy,
as already stated by Klein \&~Gr\"ave (\cite{klei86}). The brightness
peaks associated with three nebulous objects in the northern
galaxy region have mostly thermal spectra, north of the
westernmost of them a nonthermal source was found (see Table~\ref{spectral}).
The east-west elongated feature in the southern region
is also largely nonthermal, with $\alpha \simeq$ 0.6
(computed between Condon's (\cite{cond87}) and our map).

\begin{figure}
\resizebox{\hsize}{!}{\includegraphics{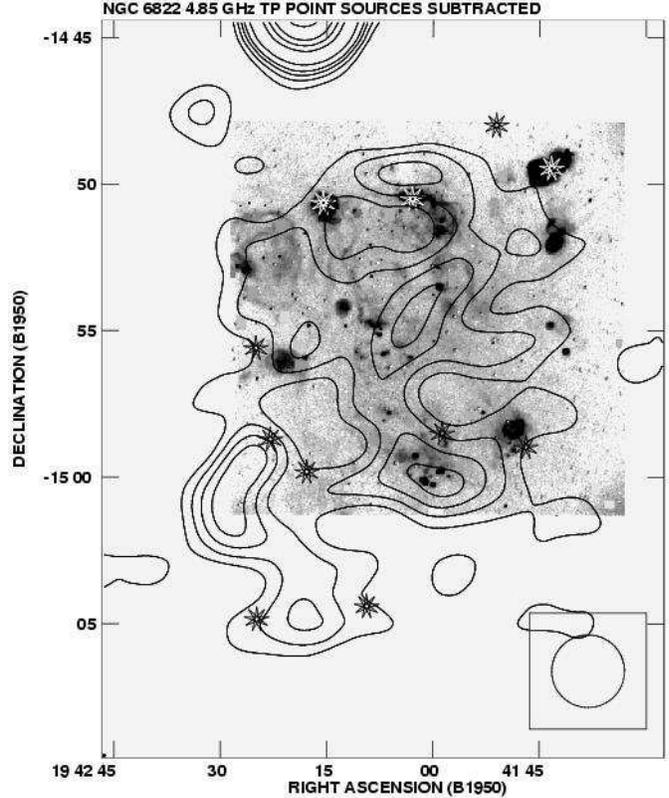}}
\caption{
The genuinely diffuse total power emission (both thermal and
nonthermal) of NGC~6822 at 4.85~GHz resulting from the
``maximum subtraction experiment'' (see text). The contour
levels are 1, 2, 3, 4, 6 etc.~mJy/b.a.. The stars show the positions of
subtracted point sources.
}
\label{subtr}
\end{figure}

In NGC~6822 we estimated the {\it minimum} content of the
genuinely diffuse total (i.e. thermal and nonthermal)
emission not clearly associated with any discrete features.
This was done by performing the ``{\it maximum} subtraction
experiment'' at the position of all sources from Condon's
(\cite{cond87}) map. We subtracted the response of the Effelsberg
telescope (taken from the calibration source map) with the
maximum amplitude not producing negative responses (thus
somewhat deeper than in the ``best subtraction'' used to
determine the spectra in Table~\ref{spectral}) allowing to estimate the
emission unexplainable by discrete sources. We did the same
at the positions of two groups of \ion{H}{ii} regions in the south-western
region at R.A.$_{1950}= 19^\mathrm{h} 41^\mathrm{m} 58\fs 7$,
Dec$_{1950}= -14\degr 58\arcmin 30\arcsec$ and R.A.$_{1950}=
19^\mathrm{h} 41^\mathrm{m} 46\fs 9$, Dec$_{1950}= -14\degr 58\arcmin
58\arcsec$ (not seen in Condon's map), to be sure that all
what remains is not associated with any discrete features.
The result is shown in Fig.\ref{subtr}.

The maximum subtraction of all possible sources of localized
emission has left a substantial diffuse component. Its
integrated flux density amounts to $59\pm 6 $~mJy, thus it
constitutes some 40\% of the total flux, the latter computed
without the strong, dominant source east of the centre. This
emission has no chance to be detected by the VLA at
frequencies $\ge$ 1.4~GHz because of its extent and low
surface brightness. In addition to a weak diffuse component
of the whole galaxy we note an extended region at
R.A.$_{1950}=19^\mathrm{h} 24^\mathrm{m} 27^\mathrm{s} $ Dec$_ {1950}=-15\degr
00\arcmin$ and its extension southwards to
R.A.$_{1950}=19^\mathrm{h} 42^\mathrm{m} 19^\mathrm{s}$ Dec$_ {1950}=-15\degr
05\arcmin$. It is coincident with the western wall of an \ion{H}{i}
shell located east of the optical galaxy, discussed by de
Blok \& Walter (\cite{debl00}). It extends southwards along the
shell, while its mentioned southernmost peak coincides with
the \ion{H}{i} maximum. The separation of this diffuse emission onto
thermal and nonthermal components has to await a reliable
determination of its spectrum and thermal content. This
requires arcmin resolution observations of NGC~6822 at
lower frequencies with an extreme sensitivity to large
smooth structures.


\begin{figure}[h]
\resizebox{\hsize}{!}{\includegraphics{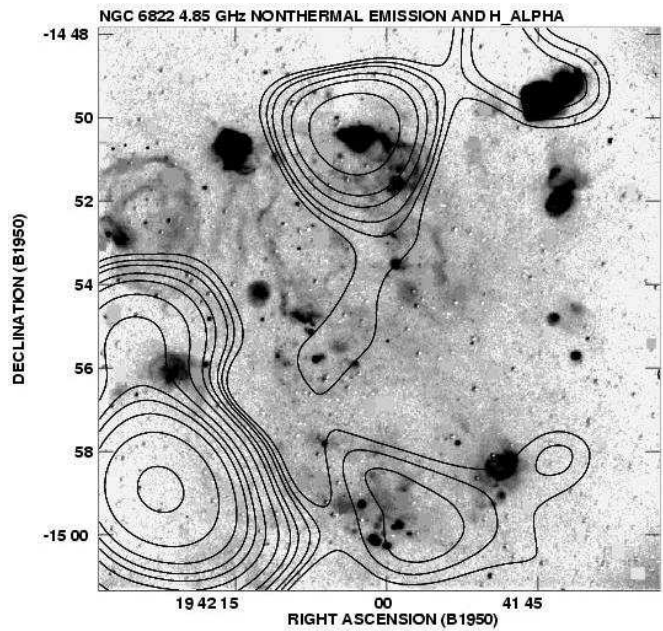}}
\caption{
The nonthermal emission at 4.85~GHz from NGC~6822 estimated
by subtracting thermal emission predicted from the H$\alpha$
emission at 4.85~GHz from our total power map. The
extinction has been calibrated using the ``best subtraction''
radio and the H$\alpha$ fluxes of three large \ion{H}{ii} regions in
the northern region of NGC~6822. The contour levels are 1,
1.5, 2, 3, 4, 6, 9, 12, 15, 20, 30, 50~mJy/b.a.
}
\label{68nt}
\end{figure}


An attempt to determine the distribution of total content of
{\it nonthermal} emission (including the discrete sources)
was made using the H$\alpha$ map. We adjusted the extinction
using the thermal sources revealed by the ``best subtraction''
method and the calibrated H$\alpha$ fluxes of associated \ion{H}{ii}
regions. NGC~6822 has a metallicity similar to that of IC 10
(Lisenfeld \& Ferrara \cite{lise98}). Thus, as the bulk of extinction
should occur in our Galaxy with little variations over the
galaxy's body, we applied the same value to a whole map. The
predicted distribution of thermal emission at 4.85~GHz was
subtracted from our total power map shown in
Fig.~\ref{68tp}. Results are shown in Fig.~\ref{68nt}. The
only definite nonthermal emission was found to come from the
eastern strong source, from the region extending northwards
of the \ion{H}{ii} complex at R.A.$_{1950}=19^\mathrm{h} 41^\mathrm{m} 44^\mathrm{s}$
Dec$_ {1950}=-19\degr 49\arcmin30\arcsec$ (seen also in
Condon's (1987) map at 1.49~GHz, and from the vicinity of
star-forming region at R.A.$_{1950}=19^\mathrm{h} 42^\mathrm{m} 02^\mathrm{s}$
Dec$_ {1950}=-19\degr 50\arcmin30\arcsec$. For the latter we
found a thermal spectrum, which signifies some nonthermal
diffuse emission around the \ion{H}{ii} region, missing in Condon's
(\cite{cond87}) map.

The evidence for the nonthermal emission from the inner
galaxy regions depends strongly on the assumption of a
constant extinction. The blob of nonthermal emission around
R.A.$_{1950}=19^\mathrm{h} 41^\mathrm{m} 56\fs$ Dec$_ {1950}=-14\degr
59\arcmin 52\arcsec$ coincides with a dense \ion{H}{i} cloud
observed by de Blok and Walter (\cite{debl00}) may be an artifact of
an underestimated H$\alpha$ flux, because of increased
extinction. A nonthermal streak extending southwards from
the \ion{H}{ii} region at R.A.$_{1950}=19^\mathrm{h} 42^\mathrm{m} 02^\mathrm{s}$
Dec$_ {1950}=-19\degr 50\arcmin30\arcsec$ may disappear if
extinction in the inner galaxy rises by some 50\% compared
to outer regions, as discussed by Massey et al. (\cite{mass95}). On 
the other hand it may get twice stronger if
the rest of the galaxy shows less extinction than the northern
star-forming regions by the same value. Even under these
favorite conditions the equipartition strength of the total
magnetic field is at most 5\,$\mu$G (assumptions like in
Sect.~\ref{ic-nth}), smaller than in the outer disk of
IC~10 and very low with respect to spiral galaxies. The
polarized emission in the northern region constitutes thus
the only unambiguous evidence for diffuse nonthermal
radiation from the inner part of NGC~6822, difficult to
detect in total power because of 6.5 times higher noise.

We also computed the mean thermal fraction in NGC~6822 by
integrating the total and thermal flux in identical areas
delineated by the boundary of the H$\alpha$ map. The thermal
emission within this area amounts to some 45\% with the bulk
of the nonthermal flux coming from the region in the SE
corner of the map shown in Fig.~\ref{68nt}. Subtracting this
region we find $\overline{f_\mathrm{th}}$ at 4.85~GHz in the inner
galaxy's body of 75\% changing between 50\% and 90\%,
depending whether the light from central part is by 50\%
more absorbed than in the outskirts (Massey at al, \cite{mass95}) or
whether there is more extinction in the northern \ion{H}{ii} regions
for which we determined directly the absorption. Moreover, a
large fraction of nonthermal emission comes from an isolated
region around the mentioned \ion{H}{ii} region in the northern part
of NGC~6822.

We estimate $\overline{f_\mathrm{th}}$ at 10.45~GHz $>$ 75\%,
likely even as high as 80\%--90\%, which is very high
compared to spiral galaxies and considerably higher than in
IC 10. Most of the observed diffuse total power emission
comes thus from the H$\alpha$ structures filling the whole
galaxy body. Week synchrotron emission with respect to the
thermal emission was also found for NGC~5907 (Dumke et al. \cite{dumk00}).

\subsection{The magnetic field structure and origin}

The only significant polarization in IC~10 has been detected
in the southern part of the galaxy, close to a straight dust
feature and to a nonthermal blob detected there by Yang \&
Skillman (\cite{yang93}). The equipartition regular field over the
rest of IC~10 is much weaker than 3\,$\mu$G. In contrast to
optically larger irregulars almost all the polarization
comes from an isolated region, while the detected regular
magnetic field is parallel to the dust lane and is probably
physically associated with local phenomena. The polarized
nonthermal blob coincides with the brightest \ion{H}{i}
concentration in IC~10 (Wilcots \& Miller \cite{wilc98}). It lies at
the southern boundary of one of large supershells blown by
winds from star-forming regions. It also lies in the galaxy
part where long \ion{H}{i} extensions with peculiar
velocities, possibly signifying gas accretion, are present.
Though no direct coincidence of our polarized feature with
these infalling gas streamers are found we speculate that it
may be a combined result of the gas kinematics
(compressions, shearing flows) driven by expanding
superbubbles and gas infall. To distinguish between genuine
regular magnetic fields and anisotropic, squeezed random
ones, produced by ``fluctuating dynamos'' (Subramanian
\cite{sub98}), then stretched and/or compressed by gas flows, needs
information on the Faraday rotation, thus multifrequency
polarization observations.

In NGC~6822 the bulk of polarized emission could be due to
background sources in the eastern galaxy and south of the
main stellar body (Fig.\ref{68pi}). On the other hand, we
note that the eastern polarized region extends along the
western wall of the \ion{H}{i} shell located east of NGC~6822 (de
Blok \& Walter \cite{debl00}), with B-vectors running almost parallel
to the wall. The southern polarized feature seems to extend
along the SW wall of the shell (as does its total power
counterpart), again with B-vectors running along the \ion{H}{i}
ridge. Moreover, it lies at the tip of the optical extension
visible e.g. in Fig.~\ref{68pi}. We cannot reject some
association of these features with local compressional
effects though the probability is rather low. High
resolution observations are required to solve this problem.

Despite its weak synchrotron emission NGC~6822 also shows
clear signs of diffuse polarized emission filling the
northern part of its optical body. The B-vectors in this
area show some alignment with the H$\alpha$ filaments
(Fig.~\ref{68tp}). The nonthermal radiation is so weak that
the polarized signal falls between 2$\sigma$ and 3$\sigma$
r.m.s. noise hence the field direction has to be taken with
extreme care. Nevertheless the polarization becomes
significant after averaging it over large areas of several
beams. Taking into account uncertainties of the thermal
fraction, NGC~6822 shows on average 5--20\% nonthermal
polarization, rising to 30--40\% in the northern half of
the galaxy. The regular field may amount to 2--3\,$\mu$G in
this region.

\begin{table*}[t]
\caption{Magnetic fields in spiral and irregular galaxies}
\label{fields}
\begin{flushleft}
\begin{tabular}{lccccl}
\hline\noalign{\smallskip}
&$<$f$_\mathrm{nt}>$ & $<$p$>$ & $<$B$_\mathrm{tot}> $&$<$B$_\mathrm{reg}>$ &
gen.\\
&10.45 GHz &nth&$\mu$G&$\mu$G&reg.\\
\hline
Normal& $\le $80\%&$\simeq $ 5\%&5--15 & 5--10
&$\alpha$--$\Omega$\\
spirals&&&&&\\
\hline
Large irr.&&&&&\\
High SF &$\simeq$60\%&$\simeq$ 5\%&5--15 & $\simeq$ 5 &
$\alpha$--$\Omega^*$\\
(NGC~4449)&&&&&\\
\hline
Small irr.&&&&&\\
High SF&$\simeq$40\%&$<5\%^1)$&5--15&$\le 3$&Fluct. dynamo+\\
(IC~10)&&$< 2\%^2)$&&&gas flows $^3)$\\
\hline
Small irr.&&&&&Magn. field\\
low SF&$\le 25\%$&$\le 20$\%&$\le 5$ &$\le 3$&diffusion? \\
(NGC~6822)&&&&&\\
\noalign{\smallskip}
\hline
\end{tabular}
\end{flushleft}
 \underline{Abbreviations:}
\newline gen./reg. -- the generating mechanism of regular
magnetic fields\\ $\alpha$--$\Omega$ -- the classical
$\alpha$--$\Omega$ dynamo described by Ruzmaikin et al.
(\cite{ruz88})\\ $\alpha$--$\Omega^*$ -- the dynamo driven by
magnetic instabilities (Moss et al. \cite{moss99}, Hanasz \& Lesch
\cite{hana00}) \\ Fluct. dynamo -- a ``fluctuating dynamo'' producing
small scale magnetic fields (Subramanian \cite{sub98})\\Magn. field
diffusion -- magnetic field produced in star-forming regions
and diffusing into the rest of the galaxy
\newline\underline {Notes:} \newline $^1)$ -- average in the
disk including compressed regions
\newline $^2)$ -- compressed regions excluded
\newline $^3)$ -- generation of random fields by the
``fluctuating dynamo'' followed by their compression and
stretching in gas flows. They will be then anisotropic
random fields rather than genuine unidirectional magnetic
fields.
\end{table*}

Table~\ref{fields} compares IC~10 and NGC~6822 to NGC~4449 -- the
irregular with much larger star-forming body (Klein et al.
\cite{klei96}, Chy\.zy et al. \cite{chyz00}) and to the mean properties of
spiral galaxies (Beck et al. \cite{beck96}, Knapik et al. \cite{knap00}). In
all cases we refer to results obtained with single dish
telescopes with a similar resolution relative to the galaxy
size. Rapidly rotating spirals are dominated by synchrotron
emission and their regular magnetic fields show typical
characteristics of the classical $\alpha-\Omega$ dynamo
process (see Beck et al. \cite{beck96}, Knapik et al. \cite{knap00}). The
irregular galaxy NGC~4449 showing weaker signs of orderly
rotation but having a quite large star-forming body
($\simeq 9$~kpc in diameter) possesses still strong regular
fields with characteristics of the dynamo process including
its strong radial magnetic field (Chy\.zy et al. \cite{chyz00},
Otmianowska-Mazur et al. \cite{otm00}).
Because of its slow rotation the classical dynamo process
may be inefficient, but its modified version driven by
Parker instabilities (Moss et al. \cite{moss99}, Hanasz \& Lesch
\cite{hana98}) may still generate strong global magnetic fields. This
galaxy has radio spectrum flatter than average for spiral
galaxies, thus probably weaker synchrotron radiation
(relative to its star-forming activity) and weaker CR
electron confinement in large-scale magnetic fields. This
may suggest a lower efficiency of the mechanism generating
global fields.

In very rapidly star-forming IC~10 having a several times
smaller optical (and ionized gas) body at least 60\% of the
flux at 10.45~GHz is of thermal origin, moreover its flat
nonthermal spectrum indicates an even smaller role of
magnetic field in the CR confinement than in NGC~4449. This
galaxy shows also much weaker global regular magnetic fields
than the larger, similarly rotating NGC~4449. Either the
mechanism producing large-scale magnetic fields is much less
efficient because of its smaller size, or a strong
concentration of star formation in a small volume destroys
regular fields efficiently. An apparent lack of a magnetized
halo, usually produced by the dynamo process, favours the
first hypothesis namely a galaxy whose size is below the
threshold of large-scale mean-field dynamo. However, random
fields may be still efficiently produced by local
``fluctuating dynamos'' (Subramanian \cite{sub98}), producing 
small-scale magnetic fields due to turbulent gas motions.

The production of a random magnetic field is even weaker
(relative to the star formation level) in NGC~6822, the
synchrotron emission fades quickly with increasing distance
from the star-forming nests. This may mean too low an energy
input from star-forming processes for the ``fluctuating
dynamo''. We note some evidence for regular fields in the
northern part of NGC~6822. It is unlikely that this is a
relic field produced during a past star formation burst, as
NGC~6822 had been quietly and slowly forming stars during
past 10~Gyr (Wyder \cite{wyder01}). We may speculate that some
magnetic field is still (or has been recently) produced by
the ``fluctuating dynamo'' in the northern star-forming
regions (note the diffuse nonthermal emission around one of
them), then diffusing into the galaxy body, getting
gradually more regular as the turbulence decays (Antonov et
al. \cite{anto00}).

\subsection{The ionized gas}

 We discovered H$\alpha$-emitting features not obviously
associated with star-forming regions in both galaxies. This
finding is surprising in NGC~6822 in which diffuse H$\alpha$
emission is found to fill the whole galaxy's body, given the
very few sources of ionising radiation.

Using our estimates of thermal and nonthermal emission we
compared the ratio of thermal and magnetic energy density
(E$_\mathrm{th}$/E$_\mathrm{B})$ in IC~10 and in the normal spiral galaxy
NGC~6946 (Ehle \& Beck \cite{ehle93}). In the latter case, for the
ionized gas properties (electron temperature, volume filling
factor) similar to those in Ehle \& Beck (\cite{ehle93})
E$_\mathrm{th}$/E$_\mathrm{B}$ was found to be between 1 and 1.5,
depending on the galactocentric radius and the exact values
of the ISM parameters. This means the energy equipartition
between the magnetic field and ionized gas. Under the same
assumptions we obtained for IC~10 values of E$_\mathrm{th}$/E$_\mathrm{B}$
ranging from 1.5--2 for a thermal fraction of 0.6 and a
thermal gas and synchrotron emission line of sight thickness
of 100~pc and 1~kpc respectively. However,
E$_\mathrm{th}$/E$_\mathrm{B}$ becomes similar to the values for NGC~6946
if either a thicker ionized gas layer (200\,pc) or thinner
synchrotron disk (800\,pc, both within uncertainties) is
adopted. Thus in IC~10 magnetic fields might play a
significant role in local gas dynamics (like the cloud
evolution, triggering star formation via MHD instabilities)
as it does in large irregulars or in large spirals.

A long filament extending to the SW and connected to \ion{H}{ii}
regions inside IC~10 seems to have comparable magnetic and
thermal energy densities (assuming a diffuse, smooth
magnetic field around the filament). The magnetic energy may
be dominant if we assume the magnetic field to be
concentrated in the filament, as indeed suggested by a
rather narrow ridge of nonthermal emission. Magnetic forces
can be important in the dynamics of this feature, but this
conclusion strongly depends on the filling factor of
magnetic field. Its knowledge needs observations with much
higher resolution.

The above is not true in the northern star-forming complex,
being completely dominated by the thermal gas. The same is
valid for a detached diffuse H$\alpha$ emitting nebulosity
west of this star-forming region. Magnetic forces seem to
play only a minor role there.

\section{Summary and conclusions}

We performed a sensitive search for the extended total power
and polarized radio emission from two irregular galaxies:
the blue compact dwarf IC~10 and the slowly star forming
NGC~6822, accompanied by sensitive mapping of the ionized
gas. The galaxies rotate slowly and are small, making the
dynamo action very difficult. The following results were
obtained:

\begin{itemize}
\item[--] IC~10 shows two isolated strongly star-forming
 nests and was found to possess a nonthermal radio
 envelope. The thermal fraction is much larger than in
 normal spirals. The galaxy shows strong but largely
 random magnetic field with a strength of 14$\pm4$\,$\mu$G.
 The nonthermal emission (signifying widespread magnetic
 fields) constitutes a smaller fraction of the total flux
 at 10.45~GHz than in spirals and in large irregulars. At
 least 60\% of emission at this frequency is likely to be
 thermal.
\item[--] NGC~6822 shows a much lower level of recent star
 formation and still possesses some weak extended radio
 continuum emission, partly of synchrotron nature which is
 difficult to prove because of lack of reliable data at
 other frequencies. Using the Mt. Laguna H$\alpha$ data
 and subtracting the bright nonthermal (probably
 background) source we find that more than 80--90\%
 of its emission (scaled to 10.45~GHz) may be thermal. The
 total magnetic field ($\le5$\,$\mu$G in the inner body) is
 very weak compared to large irregular and spiral
 galaxies.
\item[--] Both galaxies were found to possess a system of
 H$\alpha$ filaments present also outside star-forming
 regions. In IC~10 the ionized gas filament extending by
 0.68~kpc from the optically bright galaxy is associated
 with a similar synchrotron feature, suggesting a
 dynamically important magnetic field. Another, more
 clumpy ionized gas cloud is dominated by thermal
 processes.
\item[--] Both galaxies show localized traces of polarized
 emission indicating the presence of regular magnetic
 fields of strength of 2--3\,$\mu$G. In IC~10 it is
 clearly associated with a dust lane and a nonthermal
 bubble and is probably due to compression effects. In
 NGC~6822 it is much less clear, however, there is a weak
 and still disputable indication of some regular field
 associated with a large \ion{H}{i} shell. We detected also clear
 traces of diffuse polarized emission in the northern
 region of the galaxy.
\item[--] Rapidly star-forming IC~10 lacks clear signs of the
 global regular fields as expected for the classical 
 mean-field dynamo. An efficient mechanism generating purely
 random field (e.g. the ``fluctuating dynamo'', Subramanian \cite{sub98}) is
 likely to work. Nonthermal polarization seems to be
 considerably higher (locally up to 30--40\%) in the
 more quiet NGC~6822. We speculate that the magnetic field
 may be generated in scarce star-forming regions (e.g. via
 the ``fluctuating dynamo''), diffusing into quiet regions
 devoid of current star formation.

\end{itemize}

We found that the rapid rotation is not the only critical
agent in generating strong total magnetic fields filling
smoothly the whole galaxy. However, in small objects no
strong regular fields are produced though they can arise
locally due to local compressions. In case of very weak star
formation the synchrotron emission is extremely weak
compared to the thermal one. Nevertheless, even in this case
local regular fields can be found. Multifrequency radio
observations with a very good sensitivity to extended
structures are highly desirable.

\begin{acknowledgements}

The Authors wish to express their thanks to the colleagues
from the Max-Planck-Institut f\"ur Radioastronomie (MPIfR)
in Bonn for their valuable discussions during this work.
J.K., M.S., K.Ch. and M.U. are indebted to Professor Richard
Wielebinski from the MPIfR for the invitations to stay at
this Institute, where substantial parts of this work were
done. M.S. K.Ch. and M.U. are also indebted to Professor 
Ralf-J\"urgen Dettmar from Ruhr-Universit\"at Bochum for
arranging the working visits at this University. The Authors
are also grateful to Dr Marita Krause for careful reading
of the manuscript. This work was supported by a grant from
the Polish Research Committee (KBN), grants no.
962/P03/97/12 and 4264/P03/99/17. 
DJB thanks the DFG for support of the Calar observing run, H. Domg\"orgen
for participating in the October 1994 observing run, and the Humboldt
Foundation for their support from a Feodor-Lynen Fellowship during
which the observations at Mt. Laguna were taken. DJB thanks the
director and the staff of the Astronomical Observatory of the
Jagiellonian University for their hospitality during his visits.
We have made use of the LEDA database (http://leda.univ-lyon1.fr).

\end{acknowledgements}

\end{document}